\documentclass[12pt, a4paper]{article}
\usepackage{multicol} 
\usepackage{cite}
\usepackage{caption} 
\usepackage[T1]{fontenc}
\usepackage[latin9]{inputenc}
\usepackage[active]{srcltx}
\usepackage{graphicx}
\usepackage{amsmath,amsthm,amssymb,graphicx,mathtools,tikz,hyperref,tabu,enumerate,cancel,braket,float,subfigure}
\usepackage[export]{adjustbox}

\makeatletter

\newcommand{\la}{\lambda}
 
\newcommand{\vev}[1]{\langle {#1} \rangle}  
\newcommand{\order}[1]{\mathcal{O}\left({#1}\right)} 
\newcommand{\abs}[1]{\left|{#1}\right|} 
\newcommand{\Lcal}{\mathcal{L}} 
\newcommand{\met}{E_T^{\mathrm{miss}}}

\newcommand{\fbi}{\mathrm{fb}^{-1}}
\newcommand{\GeV}{\mathrm{GeV}}
\newcommand{\TeV}{\mathrm{TeV}}

\newcommand{\tchi}{\tilde{\chi}}
\newcommand{\dmc}{\Delta m_{\tilde{\chi}^\pm_1}}
\newcommand{\dmn}{\Delta m_{\tilde{\chi}^0_2}}

\newcommand{\vmet}{\vec{E}_T^\mathrm{miss}}
\newcommand{\vmpt}{\vec{p}_T^\mathrm{miss}} 
\newcommand{\vpt}{\vec{p}_T} 

\newcommand{\mhat}{\hat{\mu}}
\newcommand{\bhat}{\hat{b}}
\newcommand{\bhhat}{\hat{\hat{b}}}
\newcommand{\obs}{\mathrm{obs}}
\newcommand{\excl}{\mathrm{excl}}
\newcommand{\disc}{\mathrm{disc}}
\newcommand{\CLs}{\mathrm{CL}_s}

\@ifundefined{textcolor}{}
{%
 \definecolor{BLACK}{gray}{0}
 \definecolor{WHITE}{gray}{1}
 \definecolor{RED}{rgb}{1,0,0}
 \definecolor{GREEN}{rgb}{0,1,0}
 \definecolor{BLUE}{rgb}{0,0,1}
 \definecolor{CYAN}{cmyk}{1,0,0,0}
 \definecolor{MAGENTA}{cmyk}{0,1,0,0}
 \definecolor{YELLOW}{cmyk}{0,0,1,0}
}

\makeatother

\usepackage{babel}
\begin{document}
\begin{titlepage}
\begin{flushright}
{\tt
CTPU-PTC-21-34
}
\end{flushright}

\vskip 1.35cm 
\begin{center}

{\large{\bf 
Exploring nearly degenerate higgsinos using mono-$Z/W$ signal  
}
}

\vskip 1.5cm 

Linda M. Carpenter$^{a}$\footnote{ 
carpenter.690@osu.edu
}, 
Humberto Gilmer$^{a}$\footnote{
gilmer.30@osu.edu
}, 
Junichiro Kawamura$^{b,c}$\footnote{
jkawa@ibs.re.kr 
}

\vskip 0.8cm 

{\it 
$^a$ Department of Physics, The Ohio State University, Columbus, Ohio, 43210, USA \\ 
$^b$ Center for Theoretical Physics of the Universe, Institute for Basic Science (IBS), 
     Daejeon 34051, Korea  \\
$^c$ Department of Physics, Keio University, Yokohama, 223-8522, Japan  
}
\vspace*{0.8cm} 

\begin{abstract}
We propose a new search strategy for higgsinos.  
Assuming associated production of higgsino-like pairs with a $W$ or $Z$ boson,  
we search in the missing energy plus hadronically-tagged vector boson channel. 
We place sensitivity limits for (HL-)LHC searches 
assuming $\mathcal{O}({1\mathrm{-}3.5}~\mathrm{GeV})$ mass differences 
between the lightest neutral and charged states. 
We point out that using the $E_T^\mathrm{miss}$ distribution significantly 
increases the sensitivity of this search.  
We find the higgsinos up to 110 (210) GeV can be excluded 
with $139~(300)~\mathrm{fb}^{-1}$ data.  
The full data of the HL-LHC will exclude (discover) the higgsinos 
up to 520 (280) GeV. 
\end{abstract}
\end{center}
\end{titlepage}

\setcounter{footnote}{0}

\section{Introduction}
\label{sec-intro}

Higgsinos, supersymmetric partners of the Higgs bosons, 
are important to understand the nature 
of  electroweak (EW) symmetry breaking 
since it is closely related to the size of the EW breaking scale~\cite{Barbieri:1987fn,deCarlos:1993rbr,Chan:1997bi,Kitano:2006gv,Abe:2007kf,Barbieri:2009ev,Baer:2011ec,Papucci:2011wy,Baer:2012up,Abe:2012xm}. 
In addition, 
it is well known that the higgsino is a good candidate for dark matter (DM)~\cite{Cirelli:2005uq, Cirelli:2007xd}, 
as it has not been excluded by the direct detection of the DM~\cite{XENON:2018voc} 
if the mixing with the gauginos is sufficiently suppressed~\footnote{
See recent discussions about the higgsino DM~\cite{Co:2021ion,Chakraborti:2021kkr,Delgado:2020url,Rinchiuso:2020skh,Chen:2019gtm,Han:2019vxi,Baer:2018rhs,Kawamura:2017amp,Chun:2016cnm}.  }.
Despite these important points, 
collider limits on higgsinos remain weak due to the mass degeneracy of higgsino-like charged and neutral states; in fact the limit on the higgsino mass is about 90 GeV, obtained in the LEP experiment~\cite{ALEPH:2002gap,ALEPH:2002nwp}.
In this letter, we point out that this limit 
can be raised by using the mono-$Z/W$ boson signal at the Large Hadron Collider (LHC).

The choice of collider searches for higgsinos depends 
on the mass splitting   
between $\tchi^{\pm}_1$, $\tchi^{0}_2$ and $\tchi^0_1$, 
i.e. $\dmc := m_{\tchi^\pm_1} - m_{\tchi^0_1}$ 
and $\dmn := m_{\tchi^0_2} - m_{\tchi^0_1}$. 
Here, $\tchi^0_1$ ($\tchi^0_2$) is the (second) lightest neutral higgsino, 
and $\tchi^\pm_1$ are the charged higgsinos. 
If $\Delta m_{\chi^\pm_1} \lesssim \order{0.1~\GeV}$ 
and $\tchi^\pm_1$ is long-lived, 
the disappearing track search is available~\cite{Ibe:2006de,Mahbubani:2017gjh,Fukuda:2017jmk}.
The current limit on the pure-higgsino is 210 GeV~\cite{ATLAS:2021ttq,CMS:2020atg}. 
For larger mass splittings $\sim 0.3\mathrm{-}1~\GeV$, 
it is proposed in Ref.~\cite{Fukuda:2019kbp} that 
the higgsinos with shorter lifetime can be probed by a soft displaced track. 
On the other hand, 
soft leptons signals are available to search for higgsino pair production 
but generally require $\Delta m_{\chi^0_2} \gtrsim 5~\GeV$~\cite{Han:2014kaa,Baer:2014kya,Baer:2021srt} to probe the 100 GeV range of higgsino masses. In these searches, one hard jet from initial state radiation is required to trigger on. 
The current LHC limit is 100 (190) GeV 
for $\Delta m_{\chi^0_2} = 2~(9)~\GeV$~\cite{ATLAS:2019lng}, 
but the limit is weaker than that from the LEP experiment 
for smaller mass differences.  Thus there is a gap between the two search strategies at 
$\dmc \sim 1\mathrm{-}3.5~\GeV$ where the current limit is about 90 GeV 
given only by the LEP result.

In this letter, we point out that this gap can be probed 
by using mono-$Z/W$ signals at the LHC~\cite{CMS:2017zts,ATLAS:2018nda},  
where a EW gauge boson $V := Z,W$ decays hadronically, 
and hence is reconstructed as a large radius jet.  
A similar idea using mono-$Z$ signal is studied 
in Ref.~\cite{Anandakrishnan:2014exa}~\footnote{
The CMS searches for mono-$Z$ boson~\cite{CMS:2020ulv}. 
}, 
where the $Z$ decays leptonically.   
Since the number of signal events is small 
due to the small branching fractions of leptonic decays of a $Z$ boson, 
the limit is less than 200 GeV with the $3~\mathrm{ab}^{-1}$ data at the HL-LHC. 
We find that although backgrounds increase, 
the search with a hadronically decaying $Z/W$ boson 
may have two advantages: a larger cross section 
due to the production in associated with a $W$ boson in addition to a $Z$ boson; and a larger branching fraction to quarks. 
By 
binning the missing energy distribution, we are further able to improve the analysis. 
In fact, we find that the limit will be tightened significantly 
in the future (HL-)LHC data.

This letter proceeds as follows. 
In Sec.~\ref{sec-higgsino}, we briefly review the mass differences among the higgsinos. 
The hadronic mono-$V$ search is studied in Sec.~\ref{sec-monoV}. 
Section~\ref{sec-disc} concludes.

\section{Higgsino mass differences}
\label{sec-higgsino}

The higgsino $\tilde{H}_{u}$ ($\tilde{H}_d$) 
is the $SU(2)_L$ doublet fermion which forms a chiral supermultiplet 
with the up-type (down-type) Higgs boson $H_u$ ($H_d$). 
The mass term for the higgsinos is given by $-\mu \tilde{H}_u \tilde{H}_d$, 
where $\mu$ is the so-called $\mu$-parameter which appears in the MSSM superpotential. 
The higgsinos are mixed with the bino $\tilde{B}$ and wino $\tilde{W}$ 
which form gauge supermultiplets with the $U(1)_Y$ and $SU(2)_L$ gauge bosons, 
respectively,  after EW symmetry breaking. 
Diagonalizing the mass matrices,  
there are four neutralinos $\tchi_i^0$ ($i=1,2,3,4$) 
and two charginos $\tchi_a^\pm$ ($a=1,2$), 
where the states are ordered by increasing mass.  
We shall study the case 
that the two lighter neutralinos $\tchi_{1,2}^0$ 
and the lightest charginos $\tchi_1^\pm$ are mostly higgsino-like. 
At the tree-level~\footnote{
The mass difference induced by radiative corrections
is about $350~\mathrm{MeV}$~\cite{Nagata:2014wma}, 
so it is sub-dominant in the case of $\dmc \gtrsim 1~\GeV$, 
which is our domain of interest. 
}, 
the mass splittings of the higgsino-like states are approximately given by 
\begin{align}
 \dmn := m_{\tchi^0_2}-m_{\tchi^0_1} 
       \simeq&\  m_Z^2 \abs{\frac{c_W^2}{M_2} + \frac{s_W^2}{M_1}}, \\ 
 \dmc := m_{\tchi^\pm_1}-m_{\tchi^0_1} 
       \simeq&\  
 \frac{\dmn}{2} +  \frac{m_Z^2}{2} \sin 2\beta
           \left(\frac{s_W^2}{M_1} - \frac{c_W^2}{M_2}   \right),  
\label{eq-dmc}
\end{align}
where $M_1$ ($M_2$) is the soft mass of the bino (wino), $m_Z$ is the $Z$ boson mass, and  $c_W~(s_W):= \cos\theta_W~(\sin\theta_W)$ with $\theta_W$ is the weak mixing angle. 
The angle $\beta$ is defined as $\tan\beta := \vev{H_u}/\vev{H_d}$.   
Assuming $M_1 = M_2$, 
\begin{align}
 \dmn \sim 2 \dmc \sim 2.1~\GeV\times \left(\frac{4~\TeV}{M_2}\right),  
\end{align}
where the second term in Eq.~\eqref{eq-dmc},  
which is typically sub-dominant due to the $\tan\beta$ suppression, is neglected. 
We see that $\dmc \gtrsim 1~\GeV$ for $M_2 \lesssim 4~\TeV$, 
while $\dmn \lesssim  7~\GeV$ for $M_2 \gtrsim 1.2~\TeV$.
Hence, in a range of sub-TeV gaugino masses, 
neither is the chargino is long-lived 
nor do the neutralinos have sufficient mass gaps to leave detectable soft leptons. 
In this region of parameter space, we are thus justified in treating all the higgsino-like states $\tchi^0_{1,2}$ and $\tchi^\pm_1$ 
as invisible particles assuming moderately small mass splittings 
$\dmc \sim 1\mathrm{-}3.5~\GeV$.

\section{Mono-$Z/W$ search}
\label{sec-monoV}

\begin{figure}[t]
\centering
\begin{minipage}[t]{0.48\textwidth}
\centering
\includegraphics[height=75mm] {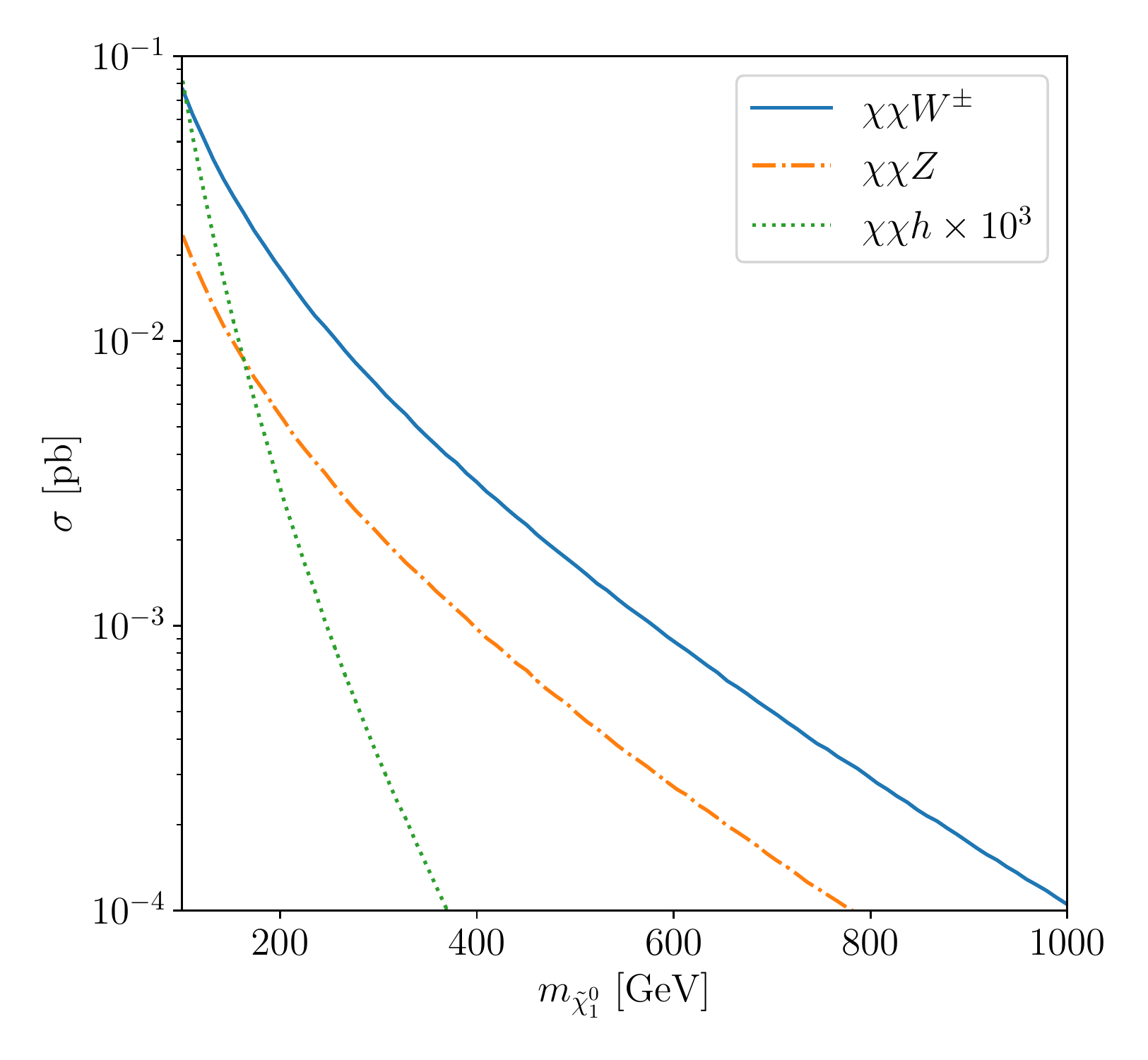}
\end{minipage}
\begin{minipage}[t]{0.48\textwidth}
\centering
\includegraphics[height=75mm] {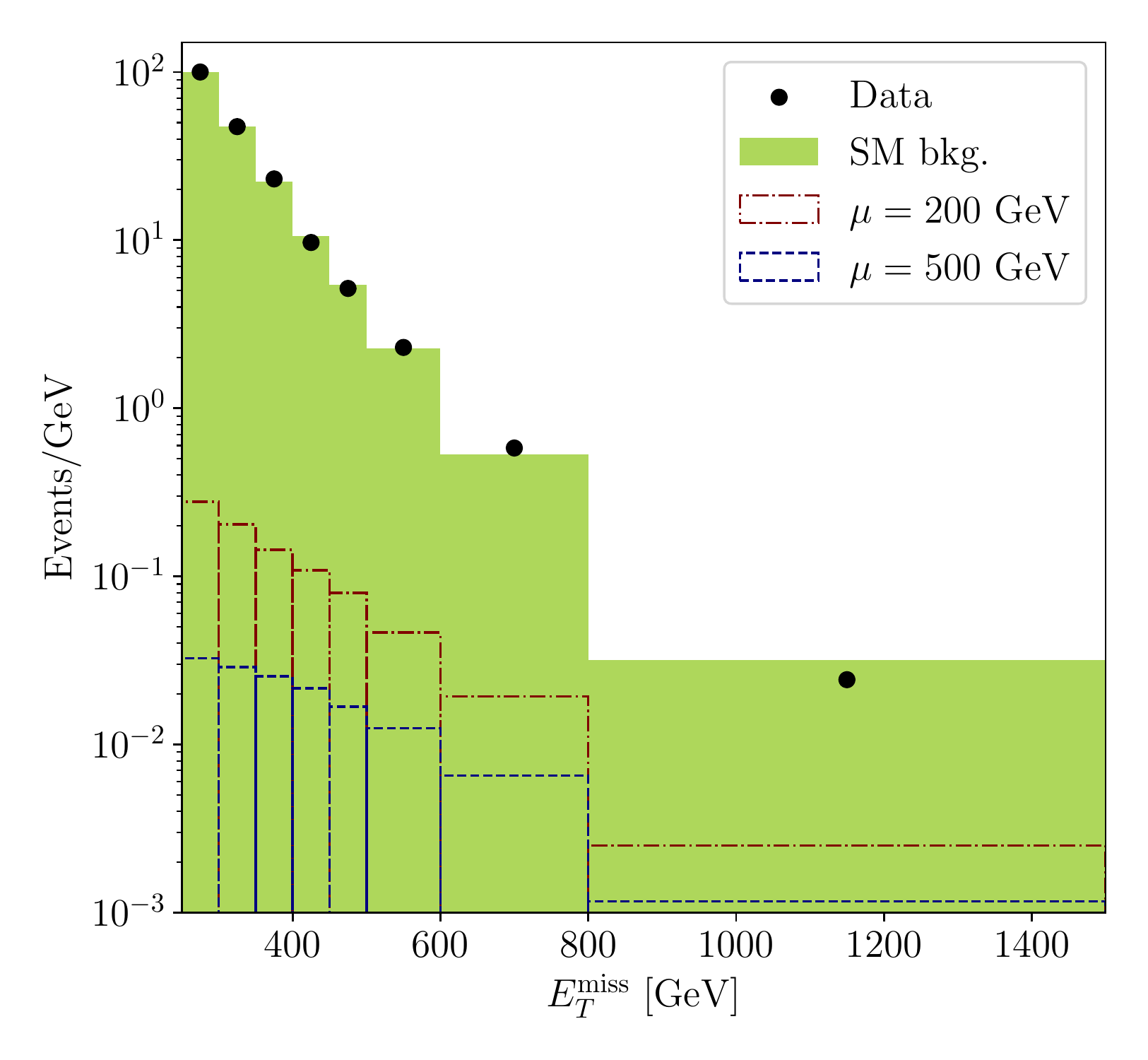}
\end{minipage}
\caption{\label{fig-xsecMET}
The production cross sections for higgsino pair production
in association with an EW gauge boson at $\sqrt{s} = 13~\TeV$ are shown on the left panel.  
The right panel shows the $\met$ distribution after the cut 
in the SR \texttt{0b-HP} when $\mu = 200$ and $500~\GeV$. 
$M_1= M_2 = 3~\TeV$ and $\tan\beta = 10$ in both panels. 
}
\end{figure}

We shall show that the mono-$V$ signals from  
 higgsino pair production in association with an EW boson $V = Z,W$, 
i.e. $pp \to \tchi \tchi V$ with $\tchi = \tchi^0_{1,2}, \tchi^\pm_1$,     
can probe higgsinos with sub-GeV mass differences in the future (HL-)LHC.  
We refer to the experimental analysis given by ATLAS~\cite{ATLAS:2018nda}. 
The left panel of Fig.~\ref{fig-xsecMET} 
shows the production cross sections of the higgsino pair productions 
in association with $W^\pm$, $Z$ bosons and the SM Higgs boson $h$ at $\sqrt{s} = 13~\TeV$ 
when $M_1=M_2 = 3~\TeV$ and $\tan\beta =10$.  
All the other soft parameters are chosen at $5~\TeV$ and $\mathrm{sgn}(\mu) = +1$.  
We shall fix these values of the soft parameters and vary $\mu$ in our analysis. 
With these parameters, $\dmc \sim 1.5~\GeV$ nearly independent of $\mu$ (for $\mu$ up to about 1 TeV). 
Cross sections are calculated by using \texttt{MadGraph-5.2.8.2}~\cite{Alwall:2014hca}
based on the sparticle spectrum calculated by \texttt{softsusy-4.1.9}~\cite{Allanach:2001kg}. 
The associated production cross section with the Higgs boson is shown for comparison, 
and it is more than $10^3$ smaller than those with the EW gauge bosons. 
We see that the associated production with a $W$ boson is dominant 
and production with a $Z$ boson is sub-dominant. 
This fact motivates us to study the hadronic decays of the EW gauge bosons 
instead of the leptonic decays of the $Z$ boson.

We simulate events 
$pp\to \tchi\tchi V(\to q\overline{q})$, with $q$ light-flavor quarks, 
using \texttt{MadGraph5}.  
The events are showered and hadronized by \texttt{Pythia8}~\cite{Sjostrand:2014zea}, 
and then run through 
the fast detector simulator \texttt{Delphes3.4.2}~\cite{deFavereau:2013fsa}. 
We used the default ATLAS card for the detector simulation, 
but we added the large-$R$ jet with $R=1.0$ on top of the small-$R$ jet with $R=0.4$
using the anti-$k_T$ jet clustering algorithm~\cite{Cacciari:2008gp,Cacciari:2011ma}. 
The trimming algorithm~\cite{Krohn:2009th} is applied 
and sub-jets with radius parameter $R=0.2$ whose transverse momenta ($p_T$) is below $5\%$ 
of the original jet $p_T$ are removed from the large-$R$ jet 
in order to remove the energy deposits from pile-up.  
We also modified 
the $p_T$ thresholds to reconstruction efficiencies of electrons and muons to be 7 GeV. 
Further, the energy fractions of the higgsino-like chargino tracks to both ECAL and HCAL are set to zero, 
the chargino tracks do not deposit energy in the calorimeters as in our parameter space they decay before encountering the tracker.

Among the signal regions (SRs) defined in Ref.~\cite{ATLAS:2018nda}, 
we find that the mono-$W/Z$ SR with 0b-tagged jet and high-purity (HP), 
namely \texttt{0b-HP}, is the most relevant for the higgsino signal. 
The SRs with b-tagged jet(s) are not very sensitive to our signal,
because the dominant production is from $\tchi \tchi W^\pm$ 
which does not have bottom quarks in the final state. 
\texttt{0b-HP} has the smallest backgrounds among the 0b-tagged jet SRs.  
In \texttt{0b-HP}, $\met > 250~\GeV$ and one large-$R$ jet is required, 
where the large $R$-jet must have $p_T > 200~\GeV$ and $\abs{\eta}<2.0$. 
Since the selection criteria for a $W/Z$ boson tagger in Ref.~\cite{ATLAS:2018nda} 
is adjusted such that the efficiency is constantly $50\%$~\cite{ATL-PHYS-PUB-2015-033}, 
we simply assume that the half of the events passing the other cuts of \texttt{0b-HP} 
are classified into the HP region.  An event is rejected if it includes any reconstructed electron (muon) 
with $p_T > 7~\GeV$ and $\abs{\eta}<2.47$ ($2.7$).  
In order to suppress the multi-jet backgrounds, 
the events are required to be 
$\Delta \phi(\vmet, \vpt^{\; J}) > 3\pi /2$, 
$\mathrm{min}_{i\in\{1,2,3\}} \Delta\phi (\vmet, \vpt^{\;j_i}) > \pi/9$, 
$\vmpt > 30~\GeV$ and $\Delta \phi(\vmet, \vmpt) < \pi/2$, 
where $\Delta\phi$ is the azimuthal angle separation between two transverse vectors. 
Here, $\vmet$, $\vmpt$, $\vpt^{\;J}$ and $\vpt^{\; j_i}$ 
are transverse momentum vectors of 
$\met$ reconstructed by \texttt{Delphes} which is calorimeter-based,  
track-based missing transverse momentum, 
large-$R$ jet and the $i$-th small-$R$ jet, respectively.  
The track-based $\vmpt$ is the negative sum of the transverse momenta of tracks 
with $p_T > 0.5~\GeV$ and $\abs{\eta} < 2.5$
except those of the charginos.
The small-$R$ jets are ordered by their $p_T$. 
Our treatment of the cuts in the other SRs 
are explained in Ref.~\cite{Carpenter:2020fnh} 
which studies limits on sneutrinos using the same signal,  
and hence we do not repeat them here.  
We found that these give weaker bounds than those from \texttt{0b-HP}.

The $\met$ distributions of the observed data, fitted SM background 
and higgsino pair production in \texttt{0b-HP} are shown 
in the right-panel of Fig.~\ref{fig-xsecMET}~\footnote{
The values in the $\met$ bins of the fitted backgrounds and errors can be found at 
\textcolor{blue}{https://www.hepdata.net/record/83180}.   
}. 
The main backgrounds are from $V+\mathrm{jets}$ production 
where an EW gauge boson decays leptonically and contributes to $\met$.  
Hence, much background is rejected by the requirement for the large-$R$ jet in the HP region, 
since the $Z/W$-tagged jet would 
be accidentally reconstructed from the multi-jet background.
This gives an advantage to the mono-$V$ search compared 
with mono-jet~\cite{CMS:2017zts,ATLAS:2021kxv} 
and mono-photon~\cite{CMS:2017qyo,ATLAS:2020uiq} searches in
which a jet (photon) is also expected in $V+\mathrm{jets}$ ($V\gamma$) backgrounds 
and are irreducibly indistinguishable from the signals. 
The red (blue) histogram shows the higgsino signal with $\mu = 200~(500)~\GeV$, 
assuming the cross section calculated by \texttt{MadGraph5} 
and the integrated luminosity of $36.1~\fbi$.
The total signal number of events passing the cuts 
is 51.0 (9.63) for $\mu = 200~(500)~\GeV$, 
and the number of events in the last bin $\met \in [800, 1500]~\GeV$ is 1.75 (0.82). 
Although the total number of events is about 5 times different at these two points, 
those in the last bin are about two times over a steeply falling background. 
Thus, a tighter cut for $\met$ will give stronger search sensitivity 
for higgsinos with heavier masses.

\begin{figure}[t]
\centering
\begin{minipage}[t]{0.49\textwidth}
\centering
\includegraphics[height=70mm] {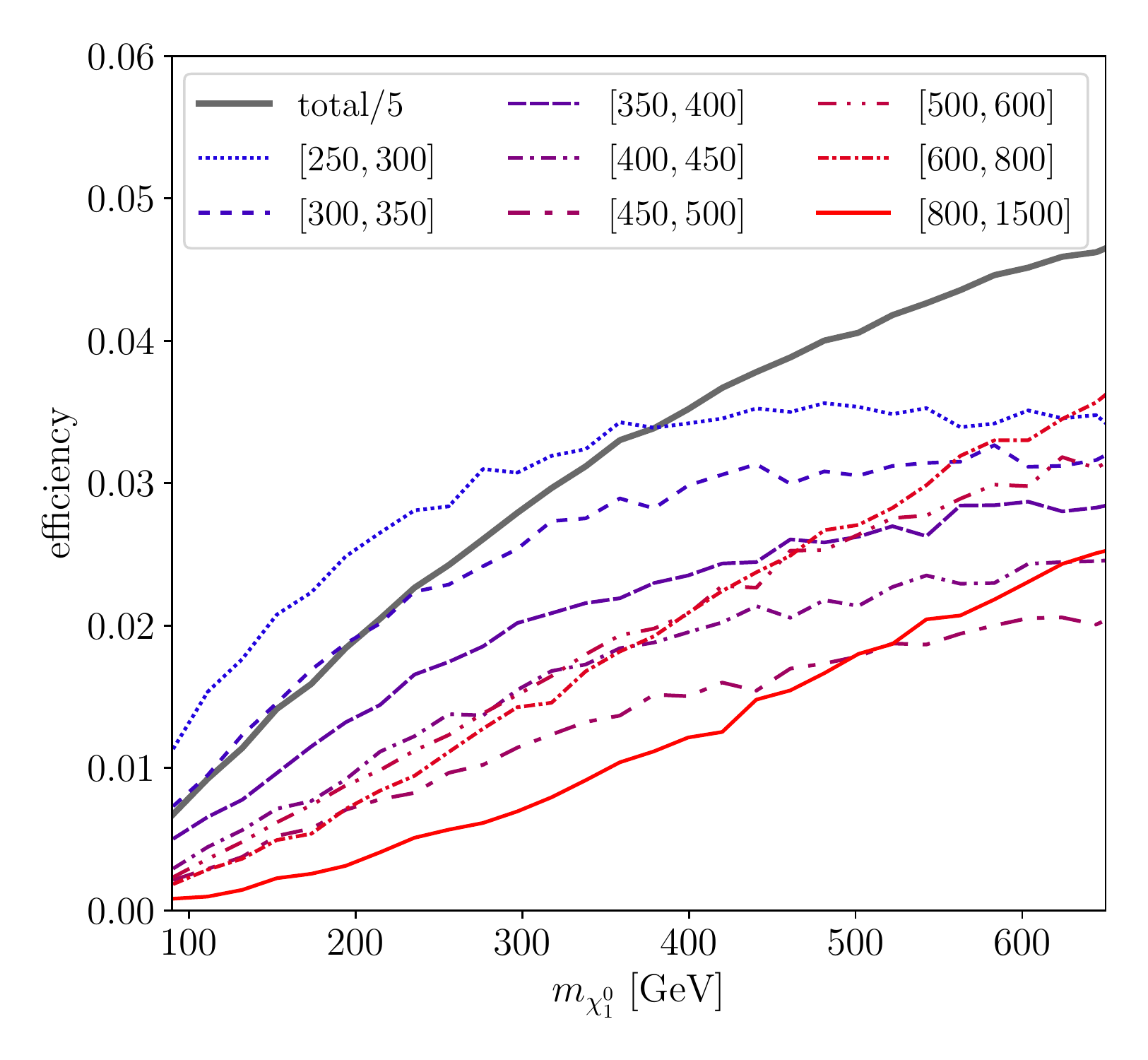}
\end{minipage}
\begin{minipage}[t]{0.49\textwidth}
\centering
\includegraphics[height=70mm] {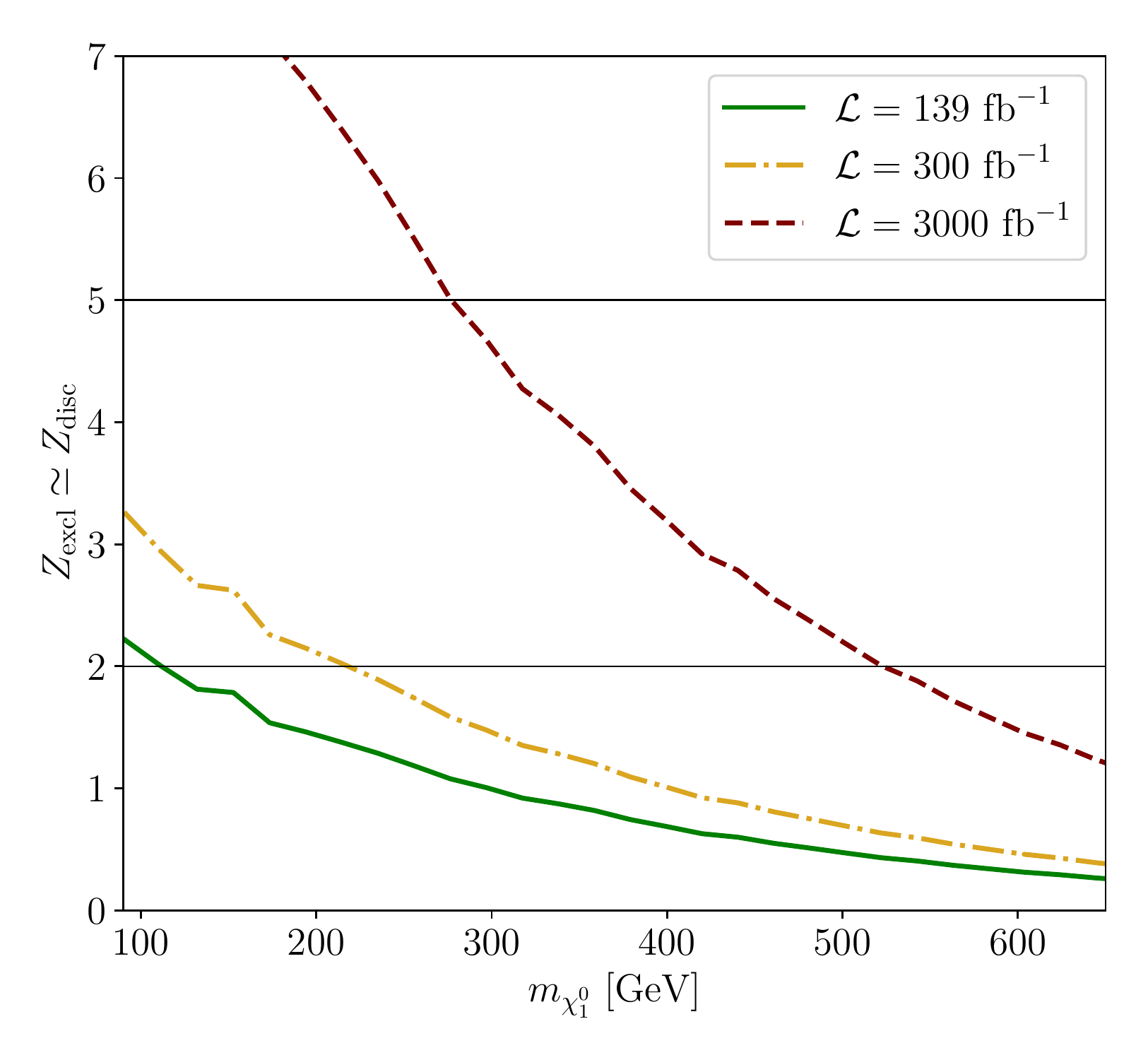}
\end{minipage}
\caption{\label{fig-limits}
The efficiencies of $\tchi\tchi V$ production of each bin of $\met$ 
in the SR \texttt{0b-HP} (left)
and the future sensitivities at the LHC (right). 
On the left panel, the efficiency to the SR \texttt{0b-HP} with $\met>250~\GeV$ 
is divided by 5 and is represented by the black line. 
}
\end{figure}

We calculate experimental limits 
based on the test statistics~\cite{CMS-NOTE-2011-005}, 
\begin{align}
 q_\mu^n := -2 \log \frac{L(n|\mu,\bhhat)}{L(n|\hat{\mu},\bhat)},  
\end{align} 
where the likelihood function is defined as 
\begin{align}
 L(n|\mu, b) := \prod_i^{N_\mathrm{bin}} \frac{\la_i^{n_i}}{n_i!}e^{-\la_i} 
                \times \frac{1}{\sqrt{2\pi} \Delta b_i} 
                \mathrm{exp}\left(-\frac{(b_i-b^0_i)^2}{2 (\Delta b_i)^2} \right), 
\end{align}
with $\la_i := \mu s_i + b_i$. 
Here, we assume a Gaussian distribution of the number of background events 
centered at $b_i^0$ with uncertainty $\Delta b_i$ in the $i$-th bin of the analysis.  
$n_i$ ($s_i$) are the number of observed (signal) events in the $i$-th bin.  
$(\mhat, \{\bhat_i\})$ is a set of values of $(\mu, \{b_i\})$ which maximizes $L$, 
and $\{\bhhat_i\}$ is that of $\{b_i\}$ which maximizes $L$ for a given $\mu$. 
Assuming the asymptotic distribution of $q_\mu$~\cite{Cowan:2010js},  
$\CLs$, $Z_\excl$ and $Z_\disc$ are respectively given by
\begin{align}
 \CLs= \frac{1-\Phi\left( \sqrt{q_1^{n_\obs}} \right)}
            {\Phi\left(\sqrt{q_1^{b_0}}-\sqrt{q_1^{n_\obs}}\right) }, 
\quad 
Z_\excl = \sqrt{q^{b_0}_1}, 
\quad \mathrm{and} \quad
Z_\disc = \sqrt{q^{s+b_0}_0},    
\end{align}
where $n_\obs$ is the number of events observed by an experiment, 
and $\Phi$ is the cumulative distribution function of the normal distribution.  
The 95\% C.L. limit is where $\CLs = 0.05$ 
and the exclusion (discovery) potential corresponds to $Z_\excl = 2$ ($Z_\disc = 5$). 
We assume that the number of background events and its error are 
simply rescaled by $R_\Lcal := \Lcal/36.1~\fbi$ and $\sqrt{R_\Lcal}$, respectively,  
with the integrated luminosities $\Lcal = 139, 300$ and $3000~\fbi$.  
These luminosities correspond to the amounts of data 
at the Run-2, Run-3 and HL-LHC respectively.

The left panel of Fig.~\ref{fig-limits}
shows efficiencies, rates to pass the cuts per a generated event, 
of the higgsino production $pp\to \tchi\tchi V$ 
in the each bin of $\met$ of \texttt{0b-HP}. 
We generated 50,000 events at each point. 
The black line represents 
the total efficiency to \texttt{0b-HP} with $\met > 250~\GeV$ divided by 5.  
We see that the efficiency to $\texttt{0b-HP}$ is about 10\% and it is dominated by the lowest $\met$ bin for the light higgsinos, 
while the total efficiency slightly increases to about 20\% 
with the higher efficiencies in the last three bins
for the heavier higgsinos. 
Since the number of signal events decreases more slowly than that of the backgrounds 
as $\met$ increases because of the efficiency increase,  
the large $\met$ bins will be able to discriminate the signal from the background 
more efficiently. 
Using the SM background (and its error) in each bin of the $\met$ distribution, 
we calculate $Z_\excl$ and $Z_\disc$.

The result of our analysis is shown in the right panel of Fig.~\ref{fig-limits}. 
We plot both $Z_\excl$ and $Z_\disc$, 
but the values are so close that we can not find visible differences.  
We see that the Run-2 (Run-3) data will put the limit about 110 (210) GeV.
However, the current data can not constrain the higgsinos heavier than the LEP bound, 
i.e. $\mathrm{CL}_s > 0.05$ for $m_{\tchi_1^0} \gtrsim 90~\GeV$.  
With the full-data of the HL-LHC, 
we can exclude (discover) higgsinos up to about 520 (280)~GeV. 
It is remarkable that the Run-2 data 
may constrain the higgsino heavier than the LEP bound. 
For comparison, if we apply the inclusive 250 GeV $\met$ cut without the $\met$ distribution binning technique can only exclude a 110 GeV higgsino with $\Lcal = 300~\fbi$ and 
the exclusion (disovery) potential is 300 (150) GeV 
with $\Lcal = 3000~\fbi$. 
Thus using the high-$\met$ bins are crucial to raise the sensitivity.

\section{Conclusion} 
\label{sec-disc}

In this letter, 
we point out that the mono-$W/Z$ search can explore the higgsinos 
with $\dmc \sim 1\mathrm{-}3.5~\GeV$, 
where the decay products of the heavier states are invisible. 
Higgsinos in this parameter range can not be covered 
by searches using disappearing tracks or soft leptons. 
Since the $\met$ distribution of the signal is different from that of the background, 
the sensitivity can be tightened by using the $\met$ distribution to obtain limits. 
We showed that the full data of the Run-2 (Run-3) at the LHC 
can exclude the higgsinos up to 110 (210) GeV, 
and that of the HL-LHC can exclude (discover) higgsinos up to about 520 (280)~GeV. 
Therefore, the mono-$W/Z$ search indeed has sensitivity to light higgsinos. 
This search could be applicable to higgsinos with smaller mass differences, 
$\dmc \lesssim 1~\GeV$ in which a charged track can be detected, 
since the search relies only on the existences of the large-$R$ jet and a large $\met$, 
and hence the signals with charged (disappearing) tracks would pass the cuts. 
Our result also suggests that 
sensitivities of searches using disappearing tracks and/or soft leptons 
could be improved 
by using a large-$R$ jet from hadronic decays of a EW gauge boson to trigger 
instead of a small-$R$ jet.

\section*{Acknowledgments}
The work of H.B.G. is supported in part by the Department of Energy (DOE) under Award No.\ DE-SC0011726.
The work of J.K.
is supported in part by
the Institute for Basic Science (IBS-R018-D1),
and the Grant-in-Aid for Scientific Research from the
Ministry of Education, Science, Sports and Culture (MEXT), Japan No.\ 18K13534.

{\small
\bibliographystyle{JHEP}
\bibliography{reference}
}

\end{document}